\documentclass[11pt,twoside]{article}  
\usepackage{asp2006}
\usepackage{adassconf}

\begin{document}   

\paperID{P4.26}

\title{Indexing astronomical database tables using HTM and HEALPix}

\markboth{Nicastro and Calderone}{Indexing astronomical DB tables using HTM
 and HEALPix}

\author{Luciano Nicastro}
\affil{INAF--IASF,
    Via P. Gobetti 101, 40129 Bologna, Italy}
\author{Giorgio Calderone}
\affil{Physics and Astronomy Department,
    Palermo University, Italy}

\contact{Luciano Nicastro}
\email{nicastro@iasfbo.inaf.it}

\paindex{Nicastro, L.}
\aindex{Calderone, G.}

\authormark{Nicastro \& Calderone}

\keywords{databases!management, databases!querying, techniques!indexing}

\begin{abstract}
In various astronomical projects it is crucial to have coordinates
indexed tables.
All sky optical and IR catalogues have up to 1 billion objects that will
increase with forthcoming projects. Also partial sky surveys at various
wavelengths can collect information (not just source lists) which can be
saved in coordinate ordered tables. Selecting a sub-set of these entries or
cross-matching them could be un-feasible if no indexing is performed.
Sky tessellation with various mapping functions have been proposed.
It is a matter of fact that the astronomical community is accepting the HTM
and HEALPix schema as the default for object catalogues and for maps
visualization and analysis, respectively.
Within the MCS library project, we have now made available as MySQL-callable
functions various HTM and HEALPix facilities. This is made possible thanks to
the capability offered by MySQL 5.1 to add external plug-ins. The DIF
(Dynamic Indexing Facilities) package distributed within the MCS library,
creates and manages a combination of Views, Triggers, DB-engine and plug-ins
allowing the user to deal with database tables indexed using one or both
these pixelisation schema in a completely transparent way.
\end{abstract}

\section{Introduction}
Spatial indexes has always been an important issue for multi dimensional data
sets in relational databases (DBs), in particular for those dealing with
spherical coordinates, e.g. latitude/longitude for Earth locations or RA/Dec for
celestial objects. Some DB servers offer built-in capabilities to
create indexes on these (coordinate) columns which consequently speed up the
execution of queries involving them. However 1. the use of these facilities
could be not easy, 2. they typically use a syntax quite different from the
astronomical one, 3. their performance is inadequate for the astronomical use.

Within the MCS library project (Calderone \& Nicastro 2007;
Nicastro \& Calderone 2006, 2007; \htmladdURL{ross.iasfbo.inaf.it/MCS/})
we have implemented the DIF package, a tool which performs and manages in a
fully
automatic way the sky pixelisation with both the HTM (Kunszt et al.\ 2001)
and HEALPix (G\'orski et al.\ 2005) schema.
Using a simple tool, any DB table with sky
coordinates columns can be easily indexed. This is achieved by using the
facilities offered by the MySQL DB server (which is the only server MCS
supports at the moment), i.e. triggers, views and plugins. Having a
table with sky coordinates, the user can make it fully
indexed in order to perform quick queries on rectangular and circular
regions (cone) or to create an HEALPix map file. An SQL query to select
objects in a cone will look like this:
\verb|SELECT * FROM MyCatalogue WHERE|
\verb|EntriesInCone(20, 30, 5)|,
where (20,30) are the coordinates of the center in degrees and 5 is the
radius in arcmin.
The important thing to note is that the DB manager needs to supply
only a few parameters in the configuration phase, whereas the generic user
does not need to know anything about the sky pixelisation either for
\texttt{SELECT} or \texttt{INSERT} or \texttt{UPDATE} queries.
It also demonstrates that there is no
need to extend standard SQL for astronomical queries (see ADQL),
at least if MySQL is used as DB server.

\section{Indexing on a sphere}
In terms of DB table indexing, mapping a sphere with a pixel scheme
means transforming a 2--d into a 1--d space, consequently a standard B--tree
index can be created on the column with the pixel IDs. On a large
astronomical table, depending on the ``depth'' of the pixelisation, this could
lead to a gain of a 4--5 orders of magnitude in search efficiency. The HTM
and HEALPix schema are widely used in Astronomy and are now well mature to
be considered as candidates for indexing tables containing astronomical data.
They are both open source
and distributed as C++ libraries. HTM uses triangular pixels which can
recursively be subdivided into four pixels. The base pixels are 8, 4 for each
hemisphere. These ``trixels'' are not equal-area but the indexing algorithm is
very efficient for selecting point sources in catalogues. HEALPix uses
equal-area pseudo-square pixels, particularly suitable for the analysis of
large-scale spatial structures. The base pixels are 12.
Using a 64 bit long integer to store the index IDs leads to a limit
for the pixels size of about 7.7 and 0.44 milli-arcsec on a side
for HTM and HEALPix, respectively.
Being able to quickly retrieve the list of objects in a given sky region is
crucial in several projects. For example hunting for transient sources like
GRBs requires fast catalogues lookup so to quickly cross match known sources
with the detected objects. The IR/optical robotic telescope REM
(Nicastro \& Calderone 2006) uses HTM
indexed catalogues to get the list of objects in $10^\prime\times 10^\prime$
regions. In this case accessing one billion objects catalogues like the
GSC2.3 takes some 10 msec.
Having a fully automatic HTM and HEALPix indexing would be crucial for the
management of the DBs of future large missions like Gaia.
Also the Virtual Observatory project would greatly benefit from adopting a
common indexing scheme for all the various types of archive it can manage.
The relevant parameters for the two pixelisations are:
\begin{center}
\begin{tabbing}
Max res. ($^{\prime\prime}$): \= $[N_{\rm pix} , 2\times N_{\rm pix}-1]$ \= ~~ \= $12 \times N_{\rm side}^2$ (where $N_{\rm side} = 2^k$) \kill

 \> \textbf{HTM}  \> \> \textbf{HEALPix}  \\
$N_{\rm pix}^\dagger$: \> $8\times 4^d$ \> \>
   $12 \times N_{\rm side}^2$ (where $N_{\rm side} = 2^k$)\\
ID range: \> $[N_{\rm pix}\;, 2\times N_{\rm pix}-1]$ \> \>
     $[0\;, N_{\rm pix}-1]$ \\
Max $N_{\rm pix}$: \> $\simeq 9.0\times 10^{15}$ \> \> $\simeq 3.5\times 10^{18}$ \\
Max res. ($^{\prime\prime}$): \> $7.7\times 10^{-3}$ \> \> $3.9\times 10^{-4}$
 ($\Omega_{\rm pix} = \pi/(3\times N_{\rm side}^2)$) \\
\rule{100pt}{0.4pt}
\\
$^\dagger d$ (depth): $[0\;, 25]$; ~
 $k$ (order $\Leftrightarrow$ resolution parameter): $[0\;, 29]$ \\
\end{tabbing}
\end{center}
As mentioned the maximum resolution is related to the usage of 64 bit integers
and it is intrinsic to the HTM and HEALPix C++ libraries.

\section{The MCS DIF package}
MCS is a set of C++ high level classes aimed at implementing
an application server, that is an application providing a service over
the network. MCS provides classes to interact with, manage and extend a MySQL
DB server. The included MyRO package allows a per row management of DB grants
whereas the DIF package allows the automatic management of sky
pixelisation with the HTM and HEALPix schema. See the
\htmladdnormallink{MCS web site}{ross.iasfbo.inaf.it/MCS/} for more information.

To enable DIF, when installing MCS it is enough to give to the configure
script the two options \verb|--enable-dif --with-mysql-source=PATH|
where \verb|PATH| is the path to the MySQL source directory. The HTM and
HEALPix C++ libraries are included in the DIF package.
A DB named \texttt{DIF} will be created containing an auxiliary table
\texttt{tbl} and a \emph{virtual} table \texttt{dif} which is dynamically
managed by the DIF DB engine.
Now let's assume one has a DB \texttt{MyDB} with a table \texttt{MyCat}
containing the two coordinates column \verb|RAcs| and \verb|DECcs|
representing the centi-arcsec
converted J2000 equatorial coordinates (this requires 4 bytes instead of the
8 necessary for a double value). To make the table manageable using both the
HTM and HEALPix pixelisation schema it is enough to give the command:
\begin{verbatim}
  dif --index-both MyDB MyCat 6 0 8 "RAcs/3.6E5" "DECcs/3.6E5"
\end{verbatim}
where \verb|dif| is the name of
the script used to perform administrative tasks related to
DIF-handled tables, 6 is the HTM depth and 8 is the HEALPix order
whereas the 0 (1) selects the RING (NESTED) scheme. The last two
parameters are the SQL expressions which convert to degrees the coordinate
values contained in the table fields \verb|RAcs| and \verb|DECcs|.
If the coordinates
where already degrees, then it would have been enough to give their names,
e.g. \verb|dif ... RA DEC|. The MySQL root password is needed.
In a future release we'll add the possibility to perform simple cross matching
between (DIF managed) catalogues.
Having an HTM indexed catalogue, the query string to obtain the list of
objects in a circular region centred on $\alpha=60^\circ$ and $\delta=30^\circ$
with radius $40^\prime$ will be:
\\
\verb|  SELECT * FROM MyCat_htm WHERE DIF_HTMCircle(60,30,40);|
\\
note the table name \verb|_htm| suffix which is needed to
actually access the view handled by DIF.
For a rectangle with the same centre and sides $50^\prime$ along the
$\alpha$ axis and $20^\prime$ along the $\delta$ axis:
\\
\verb|  SELECT * FROM MyCat_htm WHERE DIF_HTMRect(60,30,50,20);|
\\
giving only three parameters would imply a square selection.
Having chosen to use both HTM and HEALPix indexing, one could request all
the HEALPix IDs of the objects in a $50^\prime$ square by using an HTM function:
\\
\verb|  SELECT healpID FROM MyCat_htm WHERE DIF_HTMRect(60,30,50);|
\\
To simply get the IDs of the pixels falling into a circular/rectangular
region one can simply \verb|SELECT id FROM DIF.dif WHERE ...|, i.e. no
particular DIF managed table is required.
To obtain the order 10 IDs in RING scheme one can calculate them on the fly:
\\
\verb|  SELECT DIF_HEALPLookup(0,10,RAcs/3.6E5,DECcs/3.6E5)|
\\
\verb|         FROM MyCat_htm WHERE DIF_HTMCircle(60,30,20);|
\\
Giving 1 instead of 0 would give NESTED scheme IDs.
Having \verb|RA| and \verb|DEC| in degrees one would simply type
\verb|(0,10,RA,DEC)|. If one has just the HEALPix IDs
then entries on a circular region can be selected like in:
\\
  \verb|  SELECT * FROM MyCat_healp WHERE DIF_HEALPCircle(60,30,40);|
\\
note the table name \verb|_healp| suffix.
Rectangular selections for only-HEALPix indexed tables will be
available in the future. The current list of functions is: \\
\verb|DIF_HTMCircle|, \verb|DIF_HTMRect|, \verb|DIF_HTMRectV|,
\verb|DIF_HEALPCircle|, \\
\verb|DIF_HTMLookup|, \verb|DIF_HEALPLookup|, \verb|DIF_Sphedist|.
\\
\verb|DIF_HTMRectV| accepts the four corners of a
rectangle which can then have any orientation in the sky.
\verb|DIF_Sphedist| calculates the angular distance of two points on
the sphere by using the haversines formula. A first version of IDL
user contributed library and demo programs aimed at producing HEALPix
maps from the output of SQL queries is available at the
\htmladdnormallink{MCS web site}{ross.iasfbo.inaf.it/MCS/}.
\begin{figure}[t]
\epsscale{0.90}
\plotone{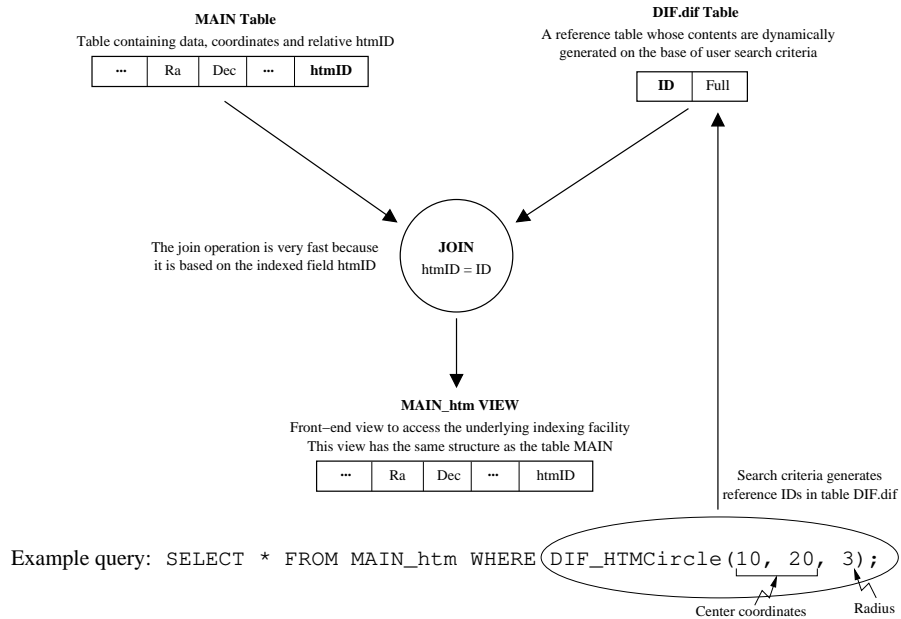}
\caption{This diagram summarises the way DIF works for the HTM indexing
case. The extra column \texttt{htmID} is added to the original table MAIN and
a view is created which performs a join with the dynamical table
\texttt{DIF.dif}. Triggers to automatically manage \texttt{INSERT} and
\texttt{UPDATE} queries are also created. We recall that MCS allows users to
interact with such MySQL server using programs written in any language.} 
\label{P4.26_fig1}
\end{figure}

\end{document}